# Continuum of consciousness: Mind uploading and resurrection of human consciousness. Is there a place for physics, neuroscience and computers?

V. Astakhov
e-mail:vadim_astakhov@hotmail.com

**Abstract**. This analysis was initiated as a joke among scholar friends but soon became interesting theoretical problem which can be mathematically formalized and even simulated in artificial environment. In this paper, I perform mental experiment to analyze hypothetical process of mind uploading. That process suggested as a way to achieve resurrection of human consciousness. Mind uploading can be seen as a migration process of the core mental functions, which migrate from a human brain to an artificial environment. Such process hypothetically might be performed through hypothetical brain-computer interface, brain transplant or prosthesis. I am looking for physical limitations which might constraint such event. To simulate the process, I suggest a topological approach which is based on a formalism of information geometry. Geometrical formalism lets us simulate the toy mind as geometrical structures as well as gives us powerful geometrical and topological methods for analysis of the information system. This approach leads to the insight about using holographic model as an analogy for the mind migration to an artificial environment. The concept of holography is well known in optics where localized 3D shape can be recorded and reconstructed by 2D dimensional hologram. I am using that analogy to develop a generalized mathematical formalism. This formalism represents the toy mind functionality in terms of information geometry (as a geometrical shape) on information manifold. Such shape can have distributed holographic representation. This representation gives us delocalized/holographic representation of the system. At the same time, the process of creating the holographic representation gives us a strategy to migrate from original to an artificial environment and back. The interface between brain and an environment is modeled as an entropy flow which is defined as a geometrical flow on information manifold. Such flow is an analogy of holography recording for the toy mind. The opposite process of holography reconstruction is modeled by none-local Hamiltonians defined on the information manifold.

## Introduction

Term "resurrection" is often discussed in religious literature but it had a limited attention from rational science. I would like to discuss this concept from the stand point of physical and bio-chemical science.

It seems reasonably to suggest that reconstruction of natural physical or biological system by assembly of their original elements does not seem feasible. First reason comes from physics, that all physical elements such as atoms and molecules are none-distinguishable. For example, any carbon atom in our body is equivalent to any other carbon atom in universe which has the same of nuclear particles and orbital electrons. Thus, question about assembly of the original elements does not make sense in context of physics.

The second argument is taken from biology and it is relevant to turnover time in biological system. Consider turnover time in biological system we can conclude that any living been is not assembled from static set of elements but rather presents continually evolving dissipative system. System components all have a finite turnover time. Most metabolites turn over within a minute in a cell, mRNA molecules typically have 2-hour half-lives in human cells, and so forth. So, a cell that you observe today, compared with the same cell yesterday, may only contain a small fraction of the same molecules. Similarly, cells have finite lifetimes. 3% of extracellular matrix in cardiac muscle is turned over daily, the cellularity of the human bone marrow turns over every 2-3 days, the renewal rate of skin is of order of 5 days to a couple of weeks, the lining of the gut epithelium has a turnover time pf about 5-7 days, and slower tissues like the liver turn over their cellularity approximately once a year. Therefore, most of the cells that are contained in an individual today were not there a few years ago. However, we consider the individual to be the same, just bit older. Likewise, we consider one cell to be the same a week later, even if most of its chemical components have turned over. Components come and go; therefore a key feature of living systems is how their components are connected together. Thus, it is reasonably to conclude that the interconnections between cells and cellular components define the essence of a living process but not specific instances of physical or chemical compounds.

Therefore, resurrection can be seen as a process which reconstructs all processes, functions and causal relations performed within the system. Why do not call it copying? Later in the paper we will provide some arguments why we treat the reconstructed system as resurrected original but not as a copy. Shortly, this is a result of the measurement process which eventually will destroy the original during any attempt to create detail physical copy. Thus, we are going to have only one instance of the system that makes the question about copy and original not relevant in context of general and quantum physics.

Now, speaking about a biological system complete reconstruction, we can get to the question "Can we resurrect lost consciousness..". This initially might seem too far from any practical applications but that can be really important in context of taking care of people in deep coma and vegetative states. Consciousness can be seen as result of complex activity of the highly organized dynamic system such as a human brain. This dynamic system exists within the rich bio-chemical environment that is provided by the bodily tissue. Tissue provides flow of mass, energy and information/entropy which are required for support of internal structure. Such system can be altered by a disease or trauma in a way which destroy its ability to generate consciousness. In that context, I would re-formulate the problem and ask another question: how rich should be the environment to provide conditions for resurrection (reconstruction or re-implementation) of the dynamic system. Later in the paper, I will provide a set of theoretical constrains to determine richness of the environment and its ability to recover lost functions. Such constraints might be suggested as a model for the tissue recovery therapies and neuroprostheses. Due to general formalism, those constraints can be applicable to other domains such as cosmological models. There might be interest to analyze how distorted physical system can be re-emerged again in the universe at letter time.

I will start with a short overview of existing theoretical models which discuss the question what resurrection might mean in terms of general physics, and what kind of universe might provide physical conditions for such an event.

"Boltzmann brain" model [1] and Tipler's cosmological model of resurrection [2] are those examples where problem of resurrection tight to the fundamental properties of our universe. The "Big Wow" model [3] (based on works of Paola Zizzi and Penrose/Hameroff [4]) suggest that early universe had conditions for consciousness to emerge ("Big Wow" state).  It is seems interesting to extend this model with Vilenkin's hypothesis of tunneling universe [5]. He suggests that early universe and physical time emerge by tunneling through initial singularity. Taking his approach and see time as a potential, I would suggest tunneling of the "Big Wow" state (alpha state in figure 1) from early universe to final singularity (beta). Tunneling will go through many trajectories, where each trajectory represents various consciousness states along the universe history. The final state beta will represent the sum over all trajectories.

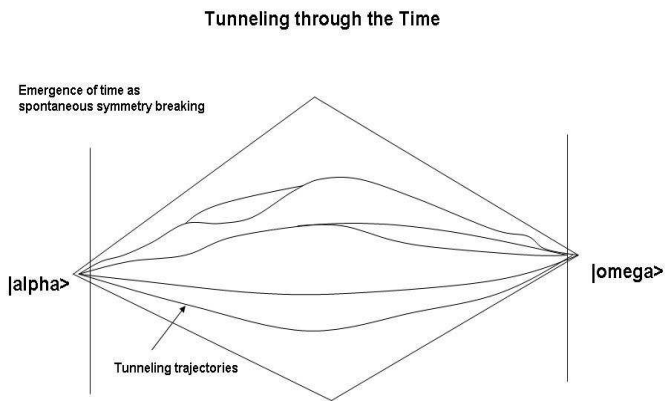

Figure 1 represent tunneling of the universe "Big Wow" state through universe time taken as a potential. Final beta state represents sum over all trajectories.

Another recently proposed theory is "big rip" model [] which is seems as another interesting candidate. The "big rip" process will reach the point where it will compete with quark gluon confinement. Ripping particle apart will require increasing amount of energy to overcome confinement. Such process might lead to bootstrap and possible entangled among wave function of those particles. That process can create a set of conditions equivalent to ones proposed by Tippler or "Big Wow" model.

Unfortunately, those scenarios do not address the biological and informational aspects of the problem. To address those issues, several hypothetical methods for mind upload might be considered: serial sectioning, tissue engineering, cloning nanotechnology and cyborning. Such consideration would immediately lead us to analysis of the whole brain mapping and the problem of copying vs. moving.

In the last decade, enthusiasts of "transhumanistic movement" claimed that human cyborgization and singularity can be achieved within this century through the development of brain-computer interfaces. These claims assume that human consciousness and mind can be migrated from native biological body to an artificial environment. In this paper, I will analyze mathematical and computational background for such hypothetical process of "mind uploading".

Following the Wikipedia definition, mind collectively refers to the aspects of intellect and consciousness manifested as combinations of thought, perception, memory, emotion, will and imagination; mind is the stream of consciousness. It seams reasonably to expect that such mind would require complex dynamic system in the background, like a brain, to operate within the reach physical environment.

Also, majority of the brain studies demonstrate that mind requires complex self-organization processes within the living brain tissue. The very basic chemical processes produce energy and building blocks for the brain neural network. Neurons and neural networks expose complex behavior such as spike generation from individual neurons, collective generation of electrical and magnetic activity as well as providing some byproducts and waist which are delivered out of the brain tissue. Thus, the migration of the mind will require us to build a rich artificial environment, which mimics the real brain complexity as well as ability for the mind to continue its vital functions.

My analysis will be increasingly focused on the systems properties of cellular networks and tissue functions. These are the properties that arise from the whole and represent physiological mechanism behind of high level cognition behaviors. These properties are sometimes referred to as "emergent" properties since they emerge from the whole and are not properties of individual parts. To simulate them, I will construct a toy model of the mind systems organization. The model will be developed as a network of causal interactions among hypothetical neural networks.

**1. Graphs and Stoichiometric matrix for an information system representation**

For simulation purposes, a toy model of the network systems is developed. It is done through networks of causal interactions among hypothetical network nodes. I suggest that the irreducible elements in the network are the elementary operation/computation which represents an elementary bio-chemical reaction or neural network activity of the hypothetical biological host. These operations can combine into function mechanisms; many functions into modules. Currently, such coarse graining of a network relies on somewhat ill-defined notions of hierarchical structure.

Figure 2 illustrates simulated system X which is a network of nodes and links. That "global network" has a "toy toy brain" $X^k_j$ which is interacting with an artificial environment X- $X^k_j$.

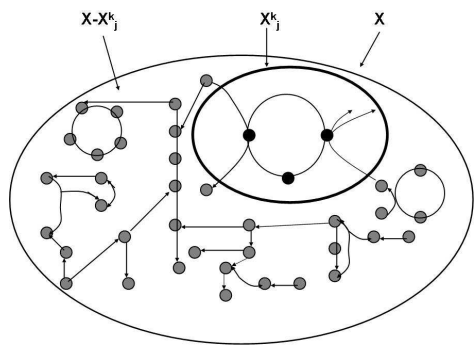

Figure 2 Left: Graph X represents the whole simulated system where $X^k_j$ is a "toy brain" interacting with an artificial environment $X-X^k_j$. Nodes are interacting compartments and links are incoming and out coming components. Right: It represents unfolding of an arbitrary node from graph X.

I am trying to keep the model as general as possible. Thus, each node depends on hierarchical level and can represent physical interaction, chemical reaction, chemical compartment, neurons or neural network. Respectively, the links can represent chemical compound participating in reaction, substances coming in and out of a compartment, neuron connections as well as connection among specialized neural networks. For examples, nodes can represent networks and links represent signal communication pathways.

It is know that neuron and neuron network functions are hierarchical and involve many layers. Therefore such high level nodes and links can be unfolded. An arbitrary neuron node can be unfolded as a network of nodes and edges representing chemical compartment and substance which coming in and leaving out the compartments. Thus, high level network functions relay on the coordinated action of the products from multiple neurons and neural networks. Such coordinated functions of multiple products can be viewed as a "causality circuit". The term causality circuit is used here to designate a collection of different neural network products and patterns that together are required to execute a particular mental function. The functions of such causal circuits are diverse, including perception, recognition, movements, decision making, and various other cognitive behaviors. Causal circuits tend to have many components; they are complex and entangled with each other. Also, they are "robust" in many but not all cases; one can remove their components without compromising their overall function. Robust collection of such entangled circuits provides background for emergence of the human mind. I am going to use term "dynamic core" to name such collection.

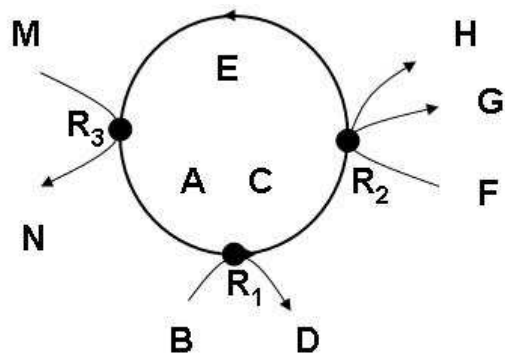

Figure 3 Illustrate simple example of a dynamic core which consists of one causality circuit. Nodes represent operations and links represent interacting compounds.

Figure 3 demonstrate an example of a simple dynamic core which consists of one causal circuit. This circuit has three function nodes in chain. Each function node operates on some external compounds and products of other functions. The nature of links between functions is complex and related to higher-order data structures. The logical consequence of this is that software function network form a tangle of cycles where different properties are being transferred through the network from one carrier to the next. This network can have many functional states.

From this standpoint, hypothetical migration process may be viewed as reconstruction of the entangled causal circuits and the whole dynamic core within certain environment. This reconstruction process will comprise four principal steps. First, the list of all components and information structures which employed by toy brain logics is enumerated. Second, the interactions between these components are studied and the "wiring diagrams" of causal circuits are constructed. Third, the causal circuits are described mathematically and their properties analyzed. Fourth, the models are used to analyze the system migration. Computer model are then generated to predict the functions that can arise from the reconstructed networks.

Functions of the reconstructed causal network will be defined by the interconnections their parts. Since these connections involve various types of interactions, they can be mathematically described by stoichiometric matrix. The matrix formalism can represents such links mathematically based on the underlying (bio-chemical) information processes. The stoichiometric matrix, which contains all relationships in a network, is thus a concise mathematical representation of reconstructed networks. This matrix comprises integers that represent time- and condition-invariant properties of a network and these properties are the key to determine the functional states of the networks that it represents.

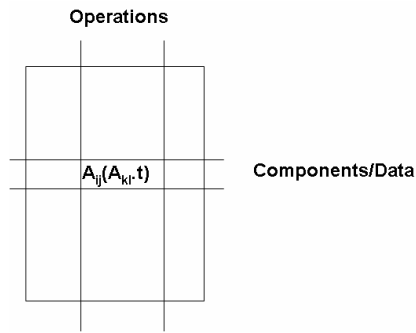

Figure 4 Stoichiometric matrixes rows represent components while columns represent operations.

All physiological processes can be simulated by information operations. The stoichiometric matrix is formed from the stoichiometric coefficients of the operations which comprise the network. The matrix is organized in such a way that every column corresponds to an operation and every row corresponds to a data structure or information object. The entries in the matrix are stoichiometric coefficients. Each row that describes an operation is constrained by the logics of this hierarchical level, for example at the bio-chemical level it will be constrained by rules of chemistry and genetics where on the level of neurons it will be rules of neuron excitation threshold, anatomy and thysiology. Such matrix transforms the information flux vector (that contains the information flow rates) into a vector that contains the time derivatives of the incoming/out coming components. The stoichiometric matrix thus contains network causal information that can be expressed by the formula: $dY/dt = S*V$, where $V = (v_1, v_2,..)$ is an information flux vector and $Y = (y_1, y_2,..)$ is a vector of the data objects. The time derivatives can be positive or negative.

After, the matrix for the whole network is constructed, one can define left null space as $L*S=0$. Thus the equation $L*S*V = d/dt(L*Y) = 0$ will provide all conservation relationships or "pools" for the system. One of such conservation relationship can define our "toy dynamic core" represented by figure 5. This figure shows stoichiometric matrix for our toy brain on figure 3. Where the columns represent causally related operations R1, R2, R3 and the rows are components which participate in the assembly of the dynamic core. "-1" means that component is an input for the operation and "1" means that component is an output of the operation.

|    | R1 | R2 | R3 |   |
|----|----|----|----|---|
|    | -1 | 0  | 1  | A |
|    | -1 | 0  | 0  | B |
|    | 1  | -1 | 0  | C |
|    | 1  | 0  | 0  | D |
|    | 0  | 1  | -1 | E |
|    | 0  | -1 | 0  | F |
|    | 0  | 1  | 0  | G |
|    | 0  | 1  | 0  | H |
|    | 0  | 0  | -1 | M |
|    | 0  | 0  | 1  | N |

Figure 5 illustrate stoichiometric matrix for the toy dynamic core on figure 2. Where the columns represent causally related operations R1, R2, R3 and the rows are components participating in the assembly of the dynamic core.

While there can be dynamic motion taking place in the column space along the operation vectors, these motions do not change the total amount of causal relations in the time-invariant dynamic core. Note that since the basis for the left null space is none unique, there are many ways to represent these pools.

The matrix formalism let us capture various types of causal relations. As a next step, I propose information geometry approach for analysis of the dynamic core migration and self assembly (self-organization) within the distributed environment. The environment consists of operation nodes which can be up and down.

## 2. Information geometry for the analysis of the network systems

Matrix formalism let us consider an arbitrary network system as a dynamic system X composed of n units $\{x_i\}$. Each unit can represent an object, an operation or a complex dynamic system. Those units can be either "on" or "off" with some probability. "On" means an element contributes to the activity that lead to emergence of the function Q and "off" is otherwise. Thus observable state of the function $Q = (Q_1, Q_2,..)$ for the system X can be characterized by certain sets of statistical parameters $(x_1, x_2, …)$ with given probability distribution $p(X|Q)$. This distribution provides causal information about elements involved in the system core dynamics.

I suggest an idea of endowing the space of such parameters with metric and geometrical structures which leads to proposal of use Fisher information as a metric of geometric space for $p(X|Q)$-distributions:

$$g_{\mu\nu} = \int (\partial p(X|Q) / \partial Q\mu)(\partial p(x|Q) / \partial Q\nu)) p(X|Q) d\{x_i\}.$$

Introduced Fisher metric is a Riemannian metric. Thus, I can define distance among states as well as other invariant functional such as affine connection $\Gamma^{\sigma}_{\lambda\nu}$, curvature tensor $R^{\lambda}_{\mu\nu k}$, Ricci tensor $R^{\mu k}$ and Curvature scalar R [7] which describes the information manifold.

$$\Gamma^{\sigma}_{\lambda\nu} = \tfrac{1}{2} g^{\sigma\nu} (\partial g_{\mu\nu} / \partial Q^{\lambda} + \partial g_{\lambda\nu} / \partial Q^{\mu} - \partial g_{\mu\lambda} / \partial Q^{\nu})$$

$$R^{\lambda}_{\mu\nu k} = \partial \Gamma^{\lambda}_{\mu\nu} / \partial Q^k - \partial \Gamma^{\lambda}_{\mu k} / \partial Q^{\nu} - \Gamma^{\eta}_{\mu\nu} \Gamma^{\lambda}_{k\eta} - \Gamma^{\eta}_{\mu k} \Gamma^{\lambda}_{\nu}$$

$$R^{\lambda}_{\mu\lambda k} = R_{\mu k}; \quad R = g^{\mu k} R_{\mu k}$$

The importance of studying information structures as geometrical structures lies in the fact that geometric structures are invariant under coordinate transforms. These transforms can be interpreted as modifications of $\{x_i\}$ set by the hardware or OS system updates with different characteristics. Thus the problem of a system survival under transition from one environment to another can be formulated geometrically.

To describe specialized sub-networks relevant to emergence of specific high level functions, I employ concept of functional cluster [8-9]: If there are any causal interactions within the system then the number of states that the system can take will be less than the number of states that its separate elements can take. Some sub-nets can strongly interact within itself and much less with other regions of the network. Geometrically, it is equivalent to higher positive curvature R~ $CI(X^k_j) = I(X^k_j) / MI(X^k_j; X - X^k_j)$ in the area of information manifold which reflects the system $X^k_j$ (figure 3) state dynamics due to the loss of information entropy "H". The loss is due to interactions among the system elements - $I(X^k_j) = \sum H(x_i) - H(X^k_j)$ and interaction with the rest of system described by mutual entropy $MI(X^k_j; X - X^k_j)$.

Curvature R near 1 indicates a subs-net which is as interactive with the rest of the system as they are within their subset. On the other hand, a R much higher than 1 will indicates the presence of a *functional cluster*- a subset of elements which are strongly interactive among themselves but only weakly interactive with the rest of the system. Localized function cluster is a manifestation of specialized region that involve in generation of high-level function. Geometrically, it will emerge as a horn area on information manifold.

Evolution of causal interactions among functional clusters is described by Ricci tensor $R_{kj}$ which is geometric analog to the concepts of *effective information* and *information integration* [9]. Effective information $EI(X^k_j \rightarrow X - X^k_j)$ between sub-net $X^k_j$ and $X - X^k_j$ can be seen as an amount of informational entropy that $X - X^k_j$ shares with $X^k_j$ due to causal effects of $X^k_j$ on $X - X^k_j$.

*2.1. Evolution of the Statistical Manifold (AdS formalism)*

Mathematically, evolution of the network system can be modeled by Euler-Lagrange equations taken from small virtual fluctuation of metric for scalar invariants. One example of such evolution equation is

$$J = -1/16\pi \int \sqrt{\det g^{\mu k}} \; (Q) \, R(Q) \, dQ.$$

External constraints can be added as a scalar term dependent on arbitrary covariant tensor $T^{\mu k}$:

$$J = -1/16\pi \int \sqrt{g(Q)} \, R(Q) \, dQ + 1/2 \int \sqrt{g(Q)} \, T^{\mu k} \, g_{\mu k} \, dQ.$$

That lead to well known geometrical equation $R^{\mu k}(Q) - g^{\mu k}(Q)R(Q) + 8\pi T^{\mu k}(Q) = 0$ which describe metric evolution. Solutions of this equation represent information systems under certain constraints defined by tensor $T^{\mu k}$. Thus, functionality of network systems can be presented in geometrical terms where AdS [10] model is one solution.

*2.2. CFT formalism*

Another way to employ geometrical approach is definition of a tangent vector space "TM(X)" for each point X of manifold M(X) as: $A_\mu \sim \partial \ln(p(x))/\partial x_\mu$ where Li brackets $[A_\mu A_\nu] \sim A_k$, give us the way to find transformations which will provide invariant descriptors. Thus, I can employ approach developed in gauge theories that are usually discussed in the language of differential geometry that make it plausible to apply for informational geometry. Mathematically, a *gauge* is just a choice of a (local) section of some principle bundle. A gauge transformation is just a transformation between two such sections. Note that although gauge theory is mainly studied by physics, the idea of a connection is not essential or central to gauge theory in general. I can define a gauge connection on the principal bundle. If a local basis of sections is chosen then it can represent covariant derivative by the connection form $A_\mu$, a Lie-algebra valued 1 –form which is called the gauge potential in physics. From this connection form it is possible to construct the curvature form *F*, a Lia-algebra valued 2-form which is an intrinsic quantity, by

$$F_{\mu\nu} = \partial_\mu A_\nu - \partial_\nu A_\mu - ig[A_\mu \, A_\nu]$$

$$[D_\mu D_\nu] = -ig \, F_{\mu\nu}, \text{ where } D_\mu = \partial_\mu - igA_\mu;$$

$$A_\mu = A_\mu^a t^a \text{ and t –is generator of infinitesimal transformation.}$$

Thus I can write Lagrangian: $1/4\ F_{\mu\nu}{}^a F_{\mu\nu}{}^a \Leftrightarrow -1/2\ Sp(F_{\mu\nu} F_{\mu\nu})$ which is invariant under transformation of coordinates. Such approach provides us with analog of CFT Yang-Mills model on our statistical manifold.

At the same time, I can have another evolution functional $J= \int (R+|\nabla f|^2)exp(-f)dV = \int (R+|\nabla f|^2)dm$, which is dependant on function f –gradient vector field defined on the manifold of volume V. It is well known in the string theory [10-13] functional that can lead us to AdS model. Thus, we have an interesting result that information geometry approach let us formulate any network system functionality in terms of two models: information analogy of super-gravity AdS model and Yang-Mills theory.

## 3. Renormalization group flow as formalism for the system migration

Now, I are going to ask a question how the evolution of the network system can be analyzed in terms of information geometry defined on information manifolds. To do so, I am looking for a functional to describe evolution of the manifold it-self.

One candidate is a functional $J=\int (R+|\nabla f|^2)dm$ which can be taken as the gradient flow $dg_{ij}/dt =2(R_{ij}+ \nabla_i \nabla_j f)$. That is generalization of the geometrical flow called Ricci flow $dg_{ij}/dt = -2R_{ij}$. Interesting thing about Ricci flow is that it can be characterized among all other evolution equations by infinitesimal behavior of the fundamental solutions of the conjugate heat equation. It is also related to the Holographic renormalization group flow [11-12] (RG-flow) which provides a structural form of the Ads-CFT correspondence.

Thus, migration of the system from one host to another can be seen as a geometrical flow of information entropy on pre-defined statistical manifold. Results [11-14] demonstrates that Ricci Flow can be considered as renormalization semi-group that distribute informational curvature over the manifold but keep scalar invariant $R=Rmin*V^{2/3}$ which is R-curvature and V-volume of the functional cluster on information manifold. As I mentioned before, the region with strong curvature interpreted as sub-system with high information integration and complexity. Thus, we can see Ricci flow as a process of delocalization which migrate functional cluster to distributed representation.

Based on Perelman works [13] for the solutions to the Ricci flow ($d/dt\ g_{ij}(t) = -2R_{ij}$) the evolution equation for the scale curvature on Riemann manifold:

$$d/dt\ R= \Delta R = 2\ |Ric|^2 = \Delta R + 2/3\ R^2\ +2|Ric^o|^2$$

It implies the estimate $R^t_{min} > - 3/(2*(t+1/4))$ where the larger t-scalar parameter then the larger is the distance scale and smaller is the energy scale. The evolution equation for the volume is $d/dt\ V < RminV$. Take R and V asymptotic at large t, we have $R(t)V(t)^{-2/3} \sim -3/2$. Thus, we have Ricci flow as a process of delocalization. If V is growing then R is decreasing and CFT model defined on the conformal boundary of the manifold. That model provides an isomorphic distributed representation for the migrated system.

In context of mind uploading and resurrection, Ricci flow provides a way to analyze the information flow within the system brain-artificial environment where functional clusters altered but the over all system functionality is preserved. This functionality can be distributed across the network of artificial nodes. The flow provides a constraint for the migration process from original toy brain system to the distributed network.

## 4. Holographic representation for information flow

Our insight of using Ricci flow based on generalization of holographic theory. Holography is a method to represent information about 3D shape by 2D record plate which called holographic plate. Expose of such 2D holographic plate by coherent light will lead to reconstruction of initial 3D shape - holography. Thus, if we have 2D record then we can reconstruct 3D shape which might be lost long time ago. This well knows optical effect has solid mathematical model called holographic representation which is based on Fourier transformation of the optical light coming from the 3D object.

Interesting fact that original 3D object is spatially localized unlike it holographic representation where information about the shape is distributed across whole 2D plate. Important to keep in mind that, an optical holography deal only with shape reconstruction and do not reconstruct the whole system which would require reconstruction all internal properties and parts as well as they interactions.

But the holographic model can be easily generalized for an abstract informational system defined on statistical manifold. The evolution of the system can be represented as a curve in the information geometry space and the whole system evolution will be represented by the set of curves in the space which called attractors. Attractor is an expression of the fact that elements form the system where they are connected and the system repeat its behavior over the time. Attractor is located in finite area of the space and presents complex shape. This shape represents a system dynamic core in terms of information geometry.

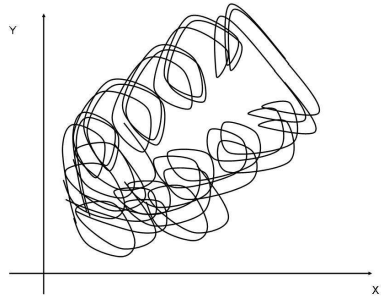

Figure 6 illustrates a 2D slice of the Dynamic core shape. The whole dynamic core of the information system is defined on multi-dimensional statistical manifold.

Based on analogy with optical 3D holography, we can construct holographic hyperspace which will encode all information about initial system attractor.

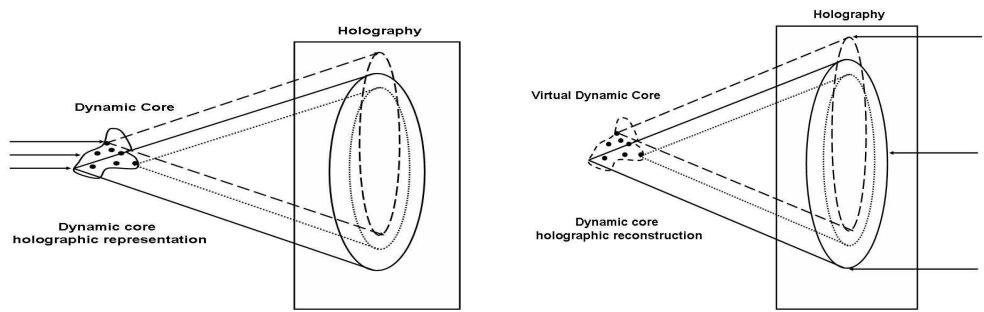

Figure 7 Figure 1 illustrates how 3D shape can be encoded (left) and decoded (right) by 2D plate. Each element of the 3D object can be encoded as a concentric curve on 2D plate. Such encoding schema allows holographic reverse engineering which will reconstruct original shape. There can be many different curves encoding each 3D point.

Figure 7 illustrate how the holographic representation on (n-1) dim plate provides delocalized representation of the original n-dim system. Information can not be localized anywhere on the (n-1) dim plate but distributed across the whole area. At the same time such holographic representation preserves all causal relations and as such the system functionality will be effectively presented by distributed environment. Holography reconstruction can be seen as a reverse engineering which will reconstruct original shape.

The question that I are going to investigate now is: how rich should be the environment to provide such conditions for holographic representation. To answer the question, a set of theoretical constrains is found. They determine richness of the environment and its ability to re-implement the toy brain functionality.

## 5. Auto poetic functionality as Coherent Structure

To estimate constraints, I am looking for the class of structures which can be re-implemented during modification in the system architecture. As it was mentioned before, deformations (geometrical flows) of the information manifold represent adjustment processes which might take place within the whole brain-computer system. Those processes should preserve functionality of the original "toy brain". In terms of geometry, vital functions should be invariant under the manifold deformations. To find out invariant structures, I investigate topological evolution of information manifold.

A function is defined by an open set of causally interacting elements which is an auto-poetic *dynamic core* of the function. Thus d*ynamic core* is defined geometrically as a coherent structure such as deformable connection domain on information manifolds with certain topological properties. I consider topological properties of such space that stay the same under continuous transformations. For the scalar function R evolves in time with some velocity VR, following the method of Cartan [15], a certain amount of topological information can be obtained by the construction of the Pfaff sequences based on the 1-form of Action, $A=ds=VR_\mu(x)dx^\mu$ a differential constructed from the unit tangent information integration velocity field VR. It is possible to claim that emergent states are coherent topological structures:

Topological Action (energy) A

Topological Vorticity (rotation) F=dA

Topological Torsion (entropy) H=A $\wedge$ dA

Topological Parity (causality) K = dA $\wedge$ dA

The rank of largest non-zero element of the above sequence gives Pfaff dimension of an information manifold. It gives us the minimal number M of functions required to determine the topological properties of the given form in a pre-geometric variety of dimensions N. It require at least dimension 4 to accommodate complex systems with dynamic cores. This Pfaff dimension is an invariant of a continuous deformation of the domain thus it is invariant under geometrical flow

It may be demonstrated from deRham theorem and Brouwer theorem [15] that the odd dimensional set (1,3,5,..) may undergo topological evolution but even dimensions (2,4,..) remain invariant. It implies that coherent topological structure once established through evolution of the Pfaff dimension from 3-to -4 then will remain invariant. Thus I have a constraint that migration of the system will be successful only if introduced changes will not alter even Pfaff dimension for the system information manifold.

## 6. Hamiltonian flow to conserve causal relations

To construct an algorithm for the toy brain migration to an artificial environment, I am using matrix representation for the system. $X_a^b$ can be seen as a matrix N*M –of N of informational objects and M operations.

Interactions among sub-systems can be expressed by an Action $S \sim m \int dt\ Tr\{X'^2_a + \omega^2[X_a\ X_b][\ X^a\ X^b]\}$ where $X_a$ is N*M matrix that can be represented X=D+Q as sum of diagonal D=diag($d_1,d_2,..$) and none-diagonal pieces. Then the action can be written as $S \sim m \int dt\ (L^D + L^Q + L^{int})$

Nelson's stochastic theory [16] emerges naturally as a description of statistical behavior of the eigenvalues with interaction potential of interaction between diagonal and none-diagonal elements:

$$L^{int} = U(D,Q)\ ;\ L^D = m \sum D'^2_a$$

$$L^Q = m\{\sum Q'^2_a + \omega^2[Q_a\ Q_b][\ Q^a\ Q^b]\}$$

$$L^{int} = 2m\ \omega^2 \sum \{-(\ d_i-d_j)^{2a}\ Q^2_b - (\ d_i-d_j)_a\ (\ d_i-d_j)_b\ Q^a\ Q^b - 2(\ d_i-d_j)^a Q^b [Q_a\ Q_b]\}$$

Based on this potential, the classical equation of motion can be written how each matrix element moves in an effective potential created by the average motion of other elements. Assume that statistical averages satisfy (Gaussians processes) relations consistent with the symmetry of the theory. That gives use Brownian movements in potential:

$$<U> = m\Omega_Q^2/2 \; Q_{ija} \; Q^{ija} + m\Omega_d^2/2 \; (d_a^i - d_a^j)^2$$

where $\Omega_Q^2 = 4(d-1)\omega^2 [(N-1)\, q^2 + 2\, r^2]$ ; $\Omega_d^2 = 4(d-1)\omega^2 \, q^2$

The Q system is in distribution. Variation principle for Matrix Model can be reformulated for eigenvalues: $\lambda = d + \sum Q_{ija}\, Q_{jia}/(d_i - d_j)a + \ldots$ . and diffusion constant for the eigenvalues is $\nu = (\Delta d)^2/\Delta t$

Now I would like to mention that neural-network can emulate quantum mechanical system at normal temperature. I are going to define $T/(8(d-1)m\, \omega^2) = t/N^p$ and $\hbar = m\, \nu$, then the wave function can be defined as: $\Psi = \sqrt{\rho} \exp(S/\hbar)$, where $\rho$ - probability density = $1/Z \exp(-H(Q)/T)$ and Hamiltonian:

$$H(Q) = m\{\sum Q'^2_a - \omega^2 [Q_a Q_b][Q^a Q^b]\}$$

This repeats Lee Smolin result that variation principle in presence of Brownian motion equivalent to the well known in math-physics Schrödinger equation:

$$i\hbar\, d\Psi/dt = \{ -\hbar/2m \; d2/d(\lambda)2 + m\Omega d2/2 \sum (\lambda_{ai} - \lambda_{aj})2 + TN(N-1)/4 + NmC\}\, \Psi$$

The Hamiltonian (and Schrodinger equation as a result of variation principle) provides an algorithm for the system migration which conserves causal circuits. Liouville's theorem (figure 8) proves that the Hamiltonian flow preserve the volume of the initial space region (representing a range of possible initial states), even though the shape of this region may become grossly distorted in the time evolution. Using Hamiltonian approach for the information geometry, thus, it can found now how the system will behave during updates described by various Hamiltonians.

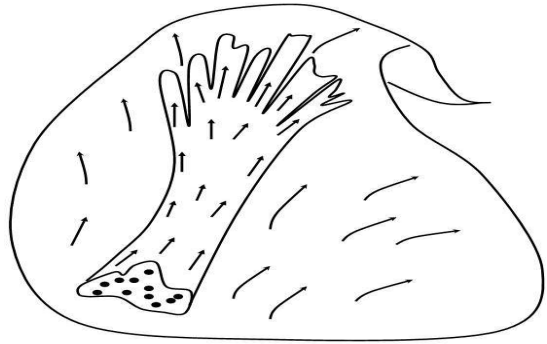

Figure 8 Liouville's theorem prove that the Hamiltonian flow preserve the volume of the initial phase-space region (representing a range of possible initial states), even though the shape of this region may become grossly distorted in the time evolution.

## 7. Migration to distributed systems by none local Hamiltonians

Here, it is possible to demonstrate how the new approach can be used in simulation. Consider again our toy brain which interacts with an environment. Each node to node interaction described by an information Hamiltonian: $H = \sum C_i * \sigma_i * \tau_i$ .

The prescribed state is denoted by the density matrix $\rho = |\psi\rangle\langle\psi|$, where $|\psi\rangle = \cos(\theta/2)|0\rangle + \exp(i\varphi)\sin(\theta/2)|1\rangle$ is superposition of active $|1\rangle$ and $|0\rangle$ none-active state. As we can see an initial state of the system of two sources is $|\rho(0)|_{4\times4}$. The information encoded in the system at time t is characterized by the fidelity $F_i(t) = \langle\psi|\rho^i(t)|\psi\rangle$, where $\rho^i(t) = \text{Tr}\,\rho(t)$-except "i".

Let us introduce two qualities:

$$CF_i(t) = \cos^2(\theta/2)\, \rho^i_{00}(t) + \sin^2(\theta/2)\, \rho^i_{11}(t)$$

$$QF_i(t) = \text{Re}[\exp(-i\varphi) * \sin(\theta)\, \rho^i_{10}(t)]$$

And rewrite Fidelity as $F_i(t) = CF_i(t) + QF_i(t)$, where $\rho(t) = \exp(-iHt)*\rho(0)*\exp(iHt)$.

By straightforward calculations [17], it is found that if $C_i = C_j$ (uniformity of information integration among all sources) for any i and j then $\sum CF_i(t)$ and $\sum QF_i(t)$ are both invariable. That mean total $F(t) = \sum F_i(t)$ is invariable too.

Due to the interactions between parts of the system, the states of these parts changes with time, and the information is expanded between them. But the total Fidelity-information is conserved. This is something like Energy Conservation Law for information systems. This law leads to interesting phenomena when information can partially concentrated spontaneously in one part of the distributed system due to oscillation part of Fi(t). Such concentration and following dilution in large-scale system can be seen as localization and de-localization of the application system Dynamic Core. That gives us another mechanism to construct algorithms for the system migration. Such systems should have "conservative Hamiltonian". Hamiltonian should satisfy condition $[H, \sum^n_{i=1} C_i] = 0$, where $C_i = 1/(2^{n-1})\, I_1 * I_2 * \ldots * \rho^0_i * I_{i+1} * I_{n+2} * \ldots * I_n$, with $\rho^0_i$ denotes the reduced density matrix of the i-th source and $I_i$ denotes reduced density matrix of other sources. This is easily can be proven by showing that $\partial F(t)/\partial t = 0$ is equal to $[H, \sum^n_{i=1} C_i] = 0$.

## 8. Computer simulation of hypothetical mind uploading and resurrection

After the mathematical formalism is developed, I am going to discuss a few cases of using hypothetical mind uploading interface. The term mind uploading (sometimes referred to by other terms such as mind transfer or whole brain emulation) refers to the transfer of a human mind to an artificial substrate. Such process will require certain physical measurements and operation. I performed mind uploading simulation study for hypothetical "toy mind" to address the list of "toy questions" such as:

Can we have complete knowledge about complex system such as a human brain?

Can we have exact copy of a physical system?

How we can construct mind uploading process?

How rich should be environment to migrate complex dynamic system?

Can we formulate "richness" in terms of mathematical and physical constraints such as "conservation lows" or "symmetries"?

Simulation Experiments did use the "toy brain" model that has a body of $10^{15}$ cells. Environment was simulated by in input parameters for the 5% of neurons. To mimic complexity of biological network, a single neuron was modeled by 5 parameters which are sensitive to measurement procedure. Then, local network circuitry was simulated for each 1000 neurons. The global neural network architecture was modeled based on the data extracted from diffusion tensor brain tomography image of the human brain. And finally, three strategies were simulated:

*8.1. Strategy 1 "Teleportation"*

"Teleportation" is the first strategy which can be considered as a candidate for mind upload through two photon teleportation. One simulated beam of entangled photons can be set to interact with the "toy brain" while another can be send to a "clone body". We simulate process of interaction which alters the states of original cells/neurons down to their molecular level. Five numeric parameters are used to simulate states of the molecular network within a cell.

We suggest that measurements of a parameter should affect $10^5$ molecular elements in the cell. This affect is modeled as an energy flux dE for each molecular structure. Taking in consideration $dxdp \sim h$ we can see that dE will be enough to change molecular thermal spectrum that will result in modification of the system parameters. For example a measurement with precision 0.1% will affect the whole system $10^5*0.1=1000\%$ by 10 times of it original state. That probably will destroy any organized systems.

Thus, the ideal physical clone is impossible, due to measurement problem.

Initially, "clone brain" was an anatomical replica of the original but without any dynamic states embodied in the molecular network of the original brain. Those states represented by dynamics of the five parameters for each neuron. Teleportation migrate those states from original to the clone. We demonstrate that cell parameters will be altered in a way that they get to uncontrolled chaotic behavior that probably would lead to cell death in the real world. Chaotic behavior came from the simulation of interactions among parameters (molecular states) and measurement device (photons).

Original will be irreversibly altered-destroyed but the clone get to the state which is indistinguishable from the state of the original before the experiment was performed. By this simulation we demonstrate that quantum "teleportation" potentially can be seen as a candidate for mind upload.

*8.2. Strategy II "Delocalization/Holographic representation"*

This strategy is based on the ideas of holography and holographic representation. The complex network interactions can be represented in terms on entropy and energy fluxes. Each interacting component of the system or an environment called "metabolite" and as such can be seen as a source of energy/entropy flux for another element of the system/environment. Thus interactions within the system/environment can be modeled as energy and entropy flows.

A statistical manifold was defined to treat "Toy brain" as a dynamic physical system. Such formalism let me analyze system dynamics in terms which are invariant to specific physical entities.

Environments can be classified as reach and poor. Reach environment is the one that provide rich structure to accommodate migration. Oppositely, the poor environment does not support such transition.

I demonstrated that system dissolution within the environment can be seen as an entropy flow. It is also demonstrated that such flow within the rich environment will conserve all causal relation of original if it satisfied renormalization group-flow constraint. Thus, such a group-flow provides holographic representation for the original host. That representation is a distributed and it provides holographic analogy for the original system. Using other words, we can say that the "mind" of our "toy brain" will survive the brain desolation and continue to run on the artificial environment that will become its new delocalized body.

*8.3. Strategy III "Resurrection"*

Finally, I consider "resurrection strategy". This is a strategy, where resurrection can be seen as a holographic reconstruction. A difference from an optical holography is that "resurrection" will destroy holographic representation due to the process of interaction. So I am still having one copy of our "toy mind".

**9. Conclusion and Future work**

In this paper, I presented an approach to analyze hypothetical process of "mind uploading" and resurrection. I demonstrated how complex network system – "toy brain" can be described in the terms of matrix theory and information geometry. Also, geometrical flow was suggested for analysis of complex dynamic structures in information geometry.

Also, Holographic representation was considered as an analogy for the "toy brain" migration. Renormalization semi-group flow and Hamiltonian flow (formalism) was taken as a way to solve the problem analytically.

Simulation study was performed and list of "toy answers" was collected:

1. I demonstrate that based on Landay principle (kTln(2) ~1bit), we can not have complete knowledge about complex system like a human brain. Measurements will require of enormous amount of thermodynamic energy to encode everything in bit.

2. Detail copy of the complex physical system is impossible due to the fact that measurements will chaotically affect the original and eventually destroy it.

3. It will be demonstrated that the processes of "information holography" (recording and reconstruction) can be seen as a resurrection, even without complete "knowledge" about the system.

4. Holography as renormalization group flow on statistical manifolds and semi-group conservation provide constraints for hypothetical processes of mind uploading from localized to delocalized form.

Next step would be a development of simulation algorithm for a large scale brain network system (~10^15 neurons).